\documentstyle[11pt,newpasp,twoside, psfig, epsfig]{article}
\def\edcomment#1{\iffalse\marginpar{\raggedright\sl#1\/}\else\relax\fi}
\reversemarginpar

\begin{document}
\title{Penetration at high-$z$ of the Greenberg `yellow stuff' : Eyes to the
Future with NGST}
\author{David L. Block}
\affil{Director: Cosmic Dust Laboratory, School of 
Computational and Applied Mathematics, University of the
Witwatersrand,
Johannesburg, South Africa}
\author{Iv\^anio Puerari}
\affil{Instituto Nacional de Astrof\'\i sica,
Optica y Electr\'onica\\
Calle Luis Enrique Erro 1, 72840 Tonantzintla, Puebla, M\'exico}
\author{Marianne Takamiya}
\affil{Department of Physics and Astronomy, University of Hawaii at Hilo\\
200 Kawili Street, Hilo, HI 96720, USA}
\author{Robert G. Abraham}
\affil{Department of Astronomy, University of Toronto\\
 60 St. George Str.,
  Toronto, ON M5S 3H8, Canada}

\begin{abstract}
Quantitative morphological dust-penetrated templates 
for galaxies in our Local Universe may also serve as excellent templates for
galaxies at high-z, because of partial/total decouplings 
expected between  gaseous and stellar
disks. 
NGC 922 is an optical irregular, which bears a striking resemblance to
objects such as HDF2-86 ($z$ = 0.749) in the HDF north (Block et al.
2001). 
Its gaseous
and stellar disk fully decouples; its stellar disk even presents
modulation of spiral arms, usually only found in grand design spiral
galaxies such as M81. Spiral galaxies
in our Local Universe appear to be open systems, that are still
forming and accreting mass, doubling their disk masses every 10 billion
years (Block et al. 2002; Bournaud \& Combes 2002). 
Likewise, galaxies at high-$z$ may also be open systems,
accreting mass, but
herein NGST will provide pivotal answers.
In this paper, we simulate the appearance of 
spiral galaxies (2--5$''$ in angular
diameter) with a class 6m Next Generation Space Telescope 
(NGST) in their dust penetrated
restframe K$'$ (2.1$\mu m$)
regime at redshifts of 0.7 and 1.2. 
Pitch angles, robustly derived from their
Fourier spectra, remain unchanged from the present to when the
Universe was roughly one half its present age.
Furthermore, a ubiquity of low m (m=1 or
m=2) spiral
wavelets or modes
is maintained in restframe K$'$ images at z=0.7 and z=1.2,
fully consistent with K$'$ morphologies for
spiral galaxies at z $\sim$ 0 in our local Universe. This paper is
dedicated to the memory of J. Mayo Greenberg, whose final research
delved into dust at high-z. Nominations for the 2004 Greenberg lecture are
invited. 
\end{abstract}

%

\section{Introduction}

Lessons learnt from studies of objects in our Local Universe is that
stellar and gaseous disks can, and do, present the most striking
decouplings
(e.g., Jarrett et al. 2003; Block et al. 1996, 1994; Block \& Wainscoat
1991). There is no
reason why such decouplings should not persist at higher redshift space.  
The evolved stellar disks of high-z galaxies in the HDF have never been
explored at restframe K$'$ band.  Images of galaxies with
redshifts ${\rm z}\sim 0.5 - 1$ or higher secured using the
Hubble Space Telescope and NICMOS never penetrate the dusty, gaseous
Population I mask. At z values greater than
3, even the H-band
1.6$\mu m$ observed flux
stems from emission shortward  of 4000 \AA. Examining
HDF galaxies at restframe I (0.84$\mu m$) would not be sufficient;
attenuation by dust even in the I band
for some local field galaxies can still be at a level of 50
percent (Block 1996).

A poignant remark pertaining to those `irregular' galaxies comprising
the faint blue excess was made by Ellis (1997):

{\it ``It is tempting
  to connect the rapidly evolving blue galaxies in the redshift
  surveys with the irregular/peculiar/merger systems ... Could this
  category of objects not simply be an increasing proportion of
  sources rendered unfamiliar by redshift or other effects? ... it is
  not yet clear whether the available local data samples are properly
  represented in all classes... the precise distinction
 between late-type spiral and
  irregular/peculiar/merger may remain uncertain.''}

In their study, Abraham et al. (1996) 
commented that the
``shape of the faint-end number count for peculiar objects is
sensitive to
the large systematic uncertainties inherent in the visual classification
of these objects''.

Submillimetre observations of galaxies at redshifts as large as 4 to 5
show that dust masses do not decrease with redshift: dust masses
at these redshifts may still be of order 10$^8$ M$_{\odot}$ (Norman
\& Braun 1996).

In this study, dedicated to J. Mayo Greenberg, we ask the
  question: What would the dust penetrated images of spiral galaxies
  at high redshift z $\sim$ 1 look like, when
the effects of dust are `swept away' and when redshift as well as
surface brightness dimming effects of $(1+$z$)^{-4}$
are fully accounted for?

The launch of the Next Generation Space Telescope (NGST) is expected
during this
decade. It is expected to be placed in an orbit
1.5 million kilometres away at the
Lagrangian point L2, which is in line with the Sun and the Earth.
The NGST science drivers dictate a
class 6m size.

Imaging capabilities are expected to
extend over the
0.6$\mu m$ to 20$\mu m$ wavelength range. Turbulence in our atmosphere
distorts wavefronts passing through, leading to rapid degradation of
the image. For NGST, any degradation of diffraction limited
performance will only be the result of distortions in the telescope
and instruments, together with pointing and tracking errors. The sky
background
will only be the result of scattered sunlight off, and emission
from, interplanetary dust grains.
The fundamental advantage of NGST for our morphological
imaging purposes
discussed here, will be the extremely low IR background shortward of
25$\mu m$.

In order to probe galaxies with redshifts 0.7 and 1.2 at restframe
K$'$,
imaging detectors in the broadband L (3.7$\mu m$; bandpass
$\delta$L=0.65$\mu m$)
and M regimes (4.7$\mu m$;
$\delta$M=0.45$\mu m$) are required.  At M, for example,
the sky background at an excellent groundbased site such as Mauna Kea
is 10 million times brighter than it will be for NGST (see Figure 1 in
Gillett \& Mountain 1998, where a sky background of $\sim$ 305
Jansky arcsec$^{-2}$
is indicated from Mauna Kea, but only 1.9$\times$10$^{-5}$ Jansky
arcsec$^{-2}$ for NGST).

Full  details of the design
of an 8192 $\times$ 8192 pixel format near-infrared camera operating
on NGST
from 0.6 to 5.3$\mu m$ (which includes the L and M bands) may be
found in Bally \&\&  Morse (1999). We follow the simulation methodology of
Takamiya (1999). The simulations recreate the images of nearby
z $\sim$ 0 spirals when moved out to higher redshifts,
always in a pre-selected restframe. Here the local spirals are simulated
at redshifts z=0.7 and z=1.2 in the dust penetrated K$'$
restframe.
Since the restframes are matched, no pixel $k$ correlations are
applied (Takamiya 1999). Furthermore, no spectral energy distribution
templates are needed: these are only required for restframes
which are not
matched, such as when simulating HDF restframe UV images
from local z $\sim$ 0
optical ones.

The angular diameters of the simulated galaxies do not
scale linearly with distance but depend on the world model adopted (see
Figure 2 in Baum 1972).
Assuming a
Friedmann-Robertson-Walker (FRW)
Big Bang cosmology with a Hubble constant of 65 km/s/Mpc and a
deceleration parameter q$_0$ of 0.1, the age of the Universe today is
approximately
14 Gyr. The lookback times for z in the interval 0.7 -- 1.2 are then
about 6.0 and 7.8 Gyr. With galaxies now confirmed to almost a
redshift of 6, galaxy formation ages shorter than 2 billion years
must be seriously considered.
Nevertheless, assuming an average epoch for galaxy
formation of 2 billion years, the evolved disks of  galaxies at z $\sim$
0.7 -- 1.2 could lie in the 4 -- 6 Gyr range.
For local field spiral galaxies observed at K$'$, the ages of `red' stars
seen in dust penetrated images can lie anywhere from 4 Gyr and older
(see the population synthesis models of Charlot et al. 1996).

Upper limits for the
lookback times at
z=0.7 and z=1.2 in a FRW cosmology where q$_0$ is increased to 0.5  are
5.5
 and 8.2
Gyr respectively; the cosmic age decreases to 10 Gyr.

Restframe UV spectra of high z galaxies are
invariably
dominated by young stellar populations only, although two exceptions
are known. They are 53W069 and 53W091 (Windhorst et al. 2000) where
Keck spectra showed that their restframe UV light was dominated by old
stellar populations with apparent ages of $\sim$ 4.5 Gyr at z=1.43 and
$\sim$ 3.5 Gyr at z=1.55.

For all our NGST simulations, we assume that the sky background is
about twice the minimum background observed by COBE (Hauser 1994).
We assume a
sky surface brightness at L and M of 19.55 and 17.31 magnitudes per square
second of arc, respectively. (From good near-infrared groundbased sites,
the corresponding sky surface backgrounds at L and M are $\sim$ 5.99
mag arcsec$^{-2}$ and --0.24 mag arcsec$^{-2}$ respectively). We
furthermore
assume pixel sizes of 50 milliarcseconds, a gain of 4 electrons per ADU
(analog digital unit), a readout noise of 4 electrons and an output
point spread
function of gaussian distribution, with a full width at half maximum
(FWHM) of 0.12$''$. Assuming eight optical surfaces with 95\%
transmission (these include the primary and
secondary mirrors as well as lenses and filters in the detector)
and a 40\% detector quantum efficiency, the
system throughput is taken to be 26\%.
For all simulations, an
on source integration time of one hour, was adopted.

The galaxies selected for our simulation runs, for which groundbased
K$'$ images are available, come from a variety of groundbased sites. 
The telescopes used for the observations
include
the 3m NASA
IRTF and the 2.2m atop Mauna
Kea, the 2.2m at La Silla and the 2.3m at Mount Stromlo.
The galaxies form part of a much larger sample
of spirals used to
develop our near-infrared classification scheme,
and full details of
each image may be found in Block et al. (2000) and references therein.

Included in Table 1 are spiral galaxies spanning the entire range of
dust-penetrated
classes, from $\alpha$ to $\gamma$. Even-sided galaxies, wherein m=2
is the dominant Fourier component, are designated by using a prefix
`E'; lopsided,
one-armed spirals (with a dominant m=1 mode) bear the `L'
prefix (Block \& Puerari 1999). Optical
arm classes of Elmegreen \& Elmegreen (1987)
are cited, when available, in column
4. The redshifts of the local sample are listed in column 5,
determined
from velocities given in the
catalogue of de Vaucouleurs et al. (1991).
When moved out to a  cosmological
distance corresponding to z=1.2, local grand design spirals such as NGC
2997 (z=0.0036) and NGC 5861 (z=0.006) only span $\sim$ 2$''$ in
diameter, whereas NGC 309,
the farthest galaxy in our local sample (z=0.0188) spans only
5$''$ in diameter. The angular diameters of all galaxies in our
simulations
lie between 2$''$ -- 5$''$.

\begin{table}
\caption{Spiral galaxies used in our pilot analysis. Column 2 gives
  the Hubble type; column 3 the dust penetrated class, column 4 the
  Elmegreen \& Elmegreen arm classes, and column 5 the redshift}
\vskip10pt
\begin{center}
\begin{tabular}{ccccc}
NGC 2857   & Sc          & E$\alpha$           & 12        & 0.0162   \\
NGC 2997   & Sc          & E$\beta$            & 9         & 0.0036   \\
NGC 309    & Sc          & E$\beta$            & 9         & 0.0188   \\
NGC 3992   & SBbc        & E$\alpha$           & 9         & 0.0035   \\
NGC 4622   & Sb          & E$\alpha$/L$\alpha$ &           & 0.0145   \\
NGC 5236   & SBc         & E$\alpha$           & 9         & 0.0019   \\
NGC 5861   & Sc          & E$\alpha$           & 12        & 0.0061   \\
NGC 7083   & Sb          & E$\gamma$           &           & 0.0102   \\
NGC 922    & Sc          & L$\gamma$           &           & 0.0102
\end{tabular}
\end{center}
\end{table}

Fourier spectra were determined for
each image at z=0.7 and z=1.2, and
inverse Fourier transform contours were then overlayed on the
simulated restframe K$'$ mosaics.
The Fourier methodology is fully described in Kalnajs (1975)
and Puerari \& Dottori (1992). The Fourier
spectra show a remarkable preservation of pitch angle in the dust penetrated
stellar disk, as a function of
redshift (see Tables 2-3). It must be stressed that the Fourier
spectra are being generated on images 2$''$ -- 5$''$ in angular size,
yet the appearance of the modes are strikingly similar to those of the
local $z\sim 0$ spectra. Dominant m=2 modes in the local sample
remain dominant at z=0.7 and z=1.2 (see column 7 in Tables 2-3), where
the dominant Fourier modes in the simulated images are given).
In fact, given two sets of Fourier
spectra, one for a  galaxy at its original distance, and another,
generated from the simulated image at 0.7 or 1.2, it is a
challenge to find any noticeable difference in their dominant
modes at all
(independent of q$_o$).
Also extremely important is the
preservation  of a ubiquity of low m (m=1 or 2) modes
present at redshifts of 0.7 and 1.2
(see Figure 3) as found in local z $\sim$ 0 galaxy images when viewed
at K$'$ (Block \& Puerari 1999; Block et al. 2000).

Our Fourier method can,
we believe, be effectively used to probe and classify evolved stellar
disks at these higher redshifts and may serve as an
excellent morphological interface between the low
and high redshift Universe.

\begin{table}
\caption{This table gives the pitch angle of every galaxy at its original
distance in column 2. Galaxies with an `S' shape have positive
values for their pitch angles, while spiral arms with `Z' shape
have negative values.
Using a 6-m NGST, columns 3 and 4 give the pitch angle in the L and
M bands, respectively, assuming q$_0$=0.1. Each pitch angle is robustly determined
from the Fourier spectra.
Columns 5 and 6 give the values
for the pitch angle at L and M for a different cosmology, wherein q$_0$=0.5.
Finally, the dominant Fourier mode in each spectrum is listed in the last column.
Note the excellent preservation of pitch angle in the restframe K$'$ images
at z=0.7 (L) and z=1.2 (M)}
\vskip10pt
\begin{center}
\begin{tabular}{ccccccc}
NGC2857 & -14.9  &   -14.0   &  -14.0    & -14.4   &  -14.4   &    (m=2) \\
NGC2997 &  25.2  &    26.5   &   25.2    &  26.5   &   26.5   &    (m=2) \\
NGC309  & -17.7  &   -17.7   &  -17.7    & -17.7   &  -17.7   &    (m=2) \\
NGC3992 &  11.0  &    10.5   &   10.5    &  11.6   &   11.8   &    (m=2) \\
NGC4622 &  -4.4  &    -4.5   &   ---     &  -4.5   &   -4.6   &    (m=1) \\
NGC4622 &   8.1  &     8.7   &   ---     &   8.7   &    8.7   &    (m=2) \\
NGC5236 & -11.9  &   -11.6   &  -11.6    & -11.6   &  -11.6   &    (m=2) \\
NGC5861 &  13.2  &    12.8   &   13.6    &  12.8   &   13.6   &    (m=2) \\
NGC7083 &  36.0  &    33.7   &   29.7    &  33.7   &   33.7   &    (m=2) \\
NGC922  & -38.6  &   -29.7   &  -29.7    & -38.6   &  -33.7   &    (m=1) \\
\end{tabular}
\end{center}
\end{table}

\null
\begin{figure}
\vspace{15truecm}
\caption{NGC 5861. Upper left: The original groundbased K$'$ image. Upper
right: Contours determined from the inverse Fourier transform are overlayed on the
groundbased K$'$ image. 
Middle left: A simulated image of NGC 5861 with an 8-m NGST
at a redshift z=0.7 (L band). Middle right: The L band image, with contours
(determined from the inverse Fourier transform) overlayed. Bottom left and right
show the galaxy redshifted to z=1.2 (M band) with and without contours.
For simulations illustrated here, a value of q$_0$=0.1 is assumed.
Notice the remarkable preservation of pitch angle with increasing
redshift. 
Also
notice the power of the Fourier method to delineate spiral arms at
z=1.2.}
\end{figure}

\begin{figure}
\vspace{15truecm}
\caption{NGC 922. An optical irregular/peculiar; a possible local
  protoype of many higher redshift `irregulars'. 
Upper left: The original groundbased K$'$ image shows a strikingly
different morphology -- even with arm modulation (Fig 3), usually only 
found in grand design spirals such as M81. 
Upper
right: Contours determined from the inverse Fourier transform are overlayed on the
groundbased K$'$ image. Middle left: A simulated image of NGC 5861 with an 8-m NGST
at a redshift z=0.7 (L band). Middle right: The L band image, with contours
(determined from the inverse Fourier transform) overlayed. Bottom left and right
show the galaxy redshifted to z=1.2 (M band) with and without contours.
For simulations illustrated here, a value of q$_0$=0.1 is assumed.
Notice the remarkable preservation of pitch angle with increasing redshift. Also
notice the power of the Fourier method to delineate spiral arms at
z=1.2.}
\end{figure}

\begin{figure}
\vspace{15truecm}
\caption{Fourier spectra may easily be generated on post-stamp images
  only 1-2 2$''$ on a side. 
The Fourier spectra of the K$'$ image of NGC 922 at its unredshifted
distance is shown at top.  The middle row shows the  Fourier spectra when the galaxy is
moved to redshifts z=0.7 (L band) and z=1.2 (M band), respectively and
a class 6m NGST. A deceleration parameter
of q$_0$=0.1 is assumed. 
The bottom row shows the  Fourier spectra for the same redshifts
values, but assuming a different cosmology with q$_0$=0.5. Notice that while there
are small changes in the shape of some of the modes such as m=1, pitch angles determined
from the dominant m=2 Fourier mode are almost identical.}
\end{figure}

\begin{figure}
\vspace{15truecm}
\caption{The template above classifies stellar disks in our Local
  Universe according to three dust penetrated arm classes ($\alpha$,
  $\beta$ and $\gamma$), linked to rotation curve shapes and rates 
of shear.  The second parameter
  is the gravitational torque of the disk. NGC 922, which might well
  serve as a local Rosetta stone for morphologically peculiar systems
  in our higher redshift universe, has a gravitational torque of two,
  in a range from 0-7, and is of class $\gamma$.}
\end{figure}

\begin{table}
\caption{As in Table 2, but for a class 8m- NGST}
\vskip10pt
\begin{center}
\begin{tabular}{ccccccc} 
NGC2857 & -14.9   &  -14.9   &  -13.6    & -14.4   &  -14.4   &    (m=2) \\
NGC2997 &  25.2   &   26.5   &   28.0    &  26.5   &   26.5   &    (m=2) \\
NGC309  & -17.7   &  -17.7   &  -17.7    & -17.7   &  -17.7   &    (m=2) \\
NGC3992 &  11.0   &   11.3   &   11.0    &  10.8   &   11.8   &    (m=2) \\
NGC4622 &  -4.4   &   -4.5   &   -4.5    &  -4.5   &   -4.5   &    (m=1) \\
NGC4622 &   8.1   &    8.4   &    8.6    &   8.4   &    8.4   &    (m=2) \\
NGC5236 & -11.9   &  -11.9   &  -11.9    & -11.9   &  -11.9   &    (m=2) \\
NGC5861 &  13.2   &   14.0   &   13.2    &  13.2   &   13.2   &    (m=2) \\
NGC7083 &  36.0   &   33.7   &   31.6    &  36.0   &   33.7   &    (m=2) \\
NGC922  & -38.6   &  -38.6   &  -45.0    & -33.7   &  -38.6   &    (m=1) \\
\end{tabular}
\end{center}
\end{table}

For local field galaxies, Block \& Puerari (1999) find a duality of
spiral structure. One classification for the Population I disk; often
a radically different one for the Population II disk. There is no
reason why this kinematical distinction should not be equally applicable
to spiral galaxies at redshifts of approximately 1. The fundamental
rather than incidental need to develop a near-infrared classification
scheme is kinematically driven; 95 percent of the mass distribution
remains
unprobed in the optical mask.  NGST will clearly cause a renaissanse in
identifying such dualities for galaxies at high redshift.

The need for an NGST larger than 4m is compelling for morphological
studies of stellar disks.
A comparative performance between a 4m NGST and 8m NGST is
evident in the simulations. Note the appearance of missing entries in
Table 4 compared to Table 2. Our
simulations always assume diffraction limited images and furthermore,
that
the pixel sizes are such that the point spread function is critically
sampled. Clearly for a 4m NGST the sampling in arcseconds is
worse than for a class 8m NGST, and the pitch angles for some of the
arms, measurable with an 8m NGST, are no longer measurable with
a class 4m NGST.
For those galaxies with a missing entry in Table 4, the
arms have become blended into the disk in the simulated 4m NGST image
and cannot be unambiguously
traced.

We have also run a series of simulations using groundbased
class 8m telescopes
equipped with L and M detectors. Due to the huge factor of
$\sim$ 10$^{7}$ Jansky arcsec$^{-2}$ corresponding to a $\sim$ 17.5
magnitude arcsec$^{-2}$
increase of the M sky background, at a redshift of 1.2 no spiral
structure
can be seen in any of the M band images. 
Thermal emission from molecules in our lower atmosphere and
from the ambient temperature of the telescope itself
dominate
the dark sky
background longward of 2.3$\mu m$ at groundbased sites.
Herein lies the great benefit of NGST. NGST is
expected
to be passively cooled to a
temperature of $\sim$ 30 K thus minimising a contribution from
the telescope itself, and of the instruments, to an unprecedently
dark sky background.

\section{Eyes to the Future: A Tribute to J. Mayo Greenberg, by DLB}

Probably the final topic which received much attention by the late
Professor J. Mayo Greenberg was the subject of dust in the
high-redshift universe: hence the dedication of this research report
to him. 

Allow one of us (DLB) some personal reflections:

When I first decided to nominate Mayo for the Henry Norris Russell
prize in 1997, letters
in support of my nomination were received by
Professors L. Spitzer (Princeton University Observatory), H. vd Hulst, G.
Miley and E. van Dishoeck (at Leiden),
R.J. Allen (Baltimore),  Bruce and Debbie Elmegreen (NY), 
D.A. Williams (University College London),
P. Hodge (Editor of the {\it Astronomical Journal} in Seattle) and L.
Allamandola, D. Cruikshank and
Y. Pendleton (NASA-Ames). The letters bore stature of the greatness of
the man. Sadly, both Lyman Spitzer and Henk vd Hulst passed on,
shortly after they wrote their recommendation letters.

As fundamental as the Hertzsprung-Russell diagram is to our
understanding of stellar evolution, so pivotal was the
pioneering insight of Mayo Greenberg to our understanding of
the
chemical composition and evolution of dust grains in our Galaxy,
and beyond.

\begin{figure}
\plotone{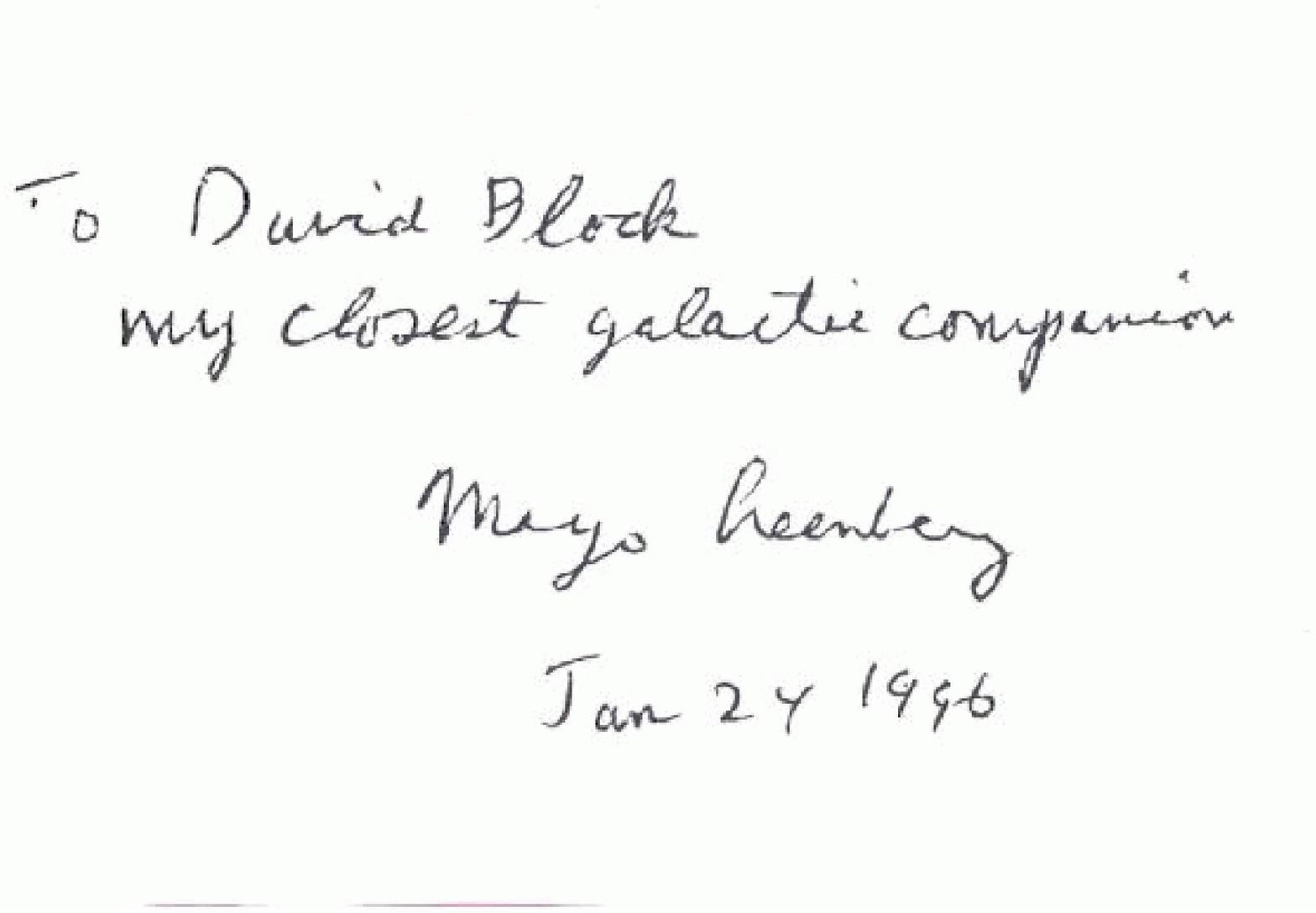}
\caption{The bond Mayo and I shared was exceptionally close. \\
Mayo
  taught me to always look up. He always encouraged me. He once
  wrote this note in my copy of his book {\it Evoluzione della polvere 
interstellare e questioni attinenti} (International School of Physics:
  Enrico Fermi; Evolution of Interstellar Dust and Related Topics).}
\end{figure}

\begin{figure}
\vspace{8truecm}
\caption{Mayo Greenberg photographed in South Africa 
with Mr Michael O'Dowd, then
  Chairman 
of the Anglo American and de Beers Chairman's Fund. 
In the background is Professor James
  Lequeux, with the author at right. }
\end{figure}

\begin{figure}
\vspace{8truecm}
\caption{Ready for a game drive in South Africa. Enjoying 
Life (capital L emphasized) to the utmost, is Mayo,
  seated in front of this 4$\times$4. Seated behind Mayo is (L-R) his wife
  Naomi;
 Liz \& Aaron Block, and Michelle Griffiths. In the
  back row
  are Ana-Maria Macchetto and Professor Richard
  Griffiths. Standing next to the 4$\times$4 is Duccio Macchetto,
  Science Director at STSCI, while a game ranger keeps guard from the
  rear of the vehicle.}
\end{figure}

\begin{figure}
\vspace{15truecm}
\caption{Mayo Greenberg 
in Pretoria, South Africa, standing on the steps of the
Voortrekker 
Monument. A contrast in profiles: one in stone, the other in
flesh. `Voor'= front in Afrikaans; `trekker' = `traveller'. 
Photograph by the author. }
\end{figure}

\begin{figure}
\vspace{8truecm}
\caption{Two master thinkers meet: Mayo Greenberg 
in deep discussion with SA Cabinet Minister Ben
  Ngubane. Copyright
  reserved by the author \& J. Waltham.}
\end{figure}

\begin{figure}
\vspace{15truecm}
\caption{Surrounded by his ever favourite `yellow stuff' in this
  caricature (by Cliff Brown) is Mayo Greenberg. All rights reserved. }
\end{figure}

\begin{figure}
\vspace{15truecm}
\caption{Mayo Greenberg in Africa. (Top) with Matthias Steinmetz
  (left), Rogier Windhorst and B. Rocca-Volmerange. Photograph by
  Cliff Brown. (Bottom) On a tour visiting Soweto. Photograph courtesy
  Liz Block.}
\end{figure}

\begin{figure}
\plotone{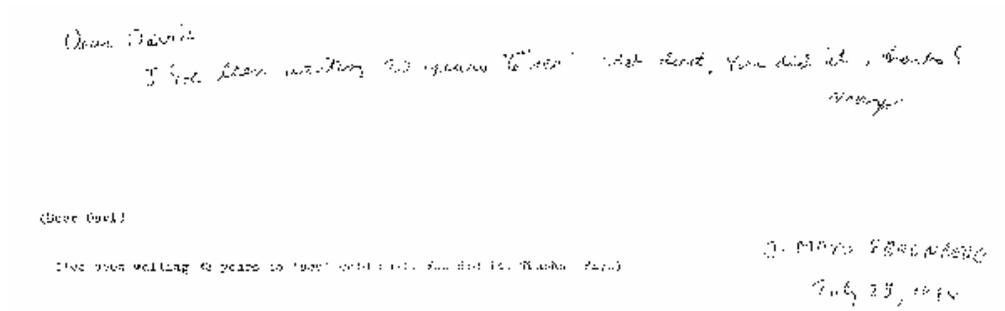}
\caption{Mayo had predicted the existence of 20K cold cosmic dust in our
  Galaxy, and beyond, decades ago. The 
observational challenge then was to find those grains which IRAS did
not detect in galaxies outside of our own.  
Mayo handed
  this note to our team of researchers at a Conference in Cardiff in 
1994: ``Iv'e been waiting 23 years to `see'
  cold dust. You did it. Thanks! Mayo.'' 
  A greater mentor cannot be envisaged.}
\end{figure}

\begin{figure}
\vspace{8truecm}
\caption{Mayo's `office' knew no bounds.  
A few quick calculations (handwritten by Mayo in a hotel room in South
Africa) appear on a 
Keck NIR spectrum of the brown dwarf Gliese 229B, published by
  B. R. Oppenheimer et al.  in 1998. The words `Oppenheimer
  et al.'  appear, in Mayo's writing, at top left. Mayo -- a man,
  heading for 80, with an unprecedented
  sweep and appreciation of current literature.}
\end{figure}

The field
in which Mayo played such an absolutely pioneering,
leading role over the decades
is now turning out to be one of the {\it most rapidly } developing areas
in galactic and extragalactic astrophysics.

Not surprising. The distribution of dust tends to delineate the location
of the
material for future generations of stars and offers evidence of a history
of past stellar processing of the ISM and metal enrichment.

As was recognised at our international conference on galaxy
morphology and cold
dust held at our University in January 1996 (attended by
over 100 astronomers worldwide), the predictive powers of Mayo's work
were absolutely remarkable. Remarkable both in depth, and in originality.

{\it He did not develop theories to explain existing observations;
  rather, Mayo developed models and predicted scores of observations}.

Conditions simulated by Professor Greenberg for the study of the
photochemistry of low temperature ices related to the photoprocessing
of interstellar dust opened up fundamentally new paths:
all of the chemical work that was done on molecules in giant
molecular clouds prior to that assumed
ion molecule chemistry, absolutely ignoring the fact that there was
dust there. Researchers had ignored it from so many points of view,
it would be impossible for me to enumerate them.

Greenberg's far-reaching insights into
the pivotal
role which the evolution of dust grains play in  our understanding of
star formation, are only now being fully appreciated.
His now famous silicate core/organic refractory
and ice mantle model for dust grains was developed {\it decades} before
the advent of modern ground and space based instruments could test the
models.

I think in particular, of the confirmation from spectroscopic studies
of the silicate spectral features at 9.7$\mu m$ and 18$\mu m$ and the
3.4 $\mu m$ organic feature in the diffuse interstellar medium.

But there is more. Dust is swept by stellar winds and
explosions into clumps, clouds, shells, and filaments that can collapse to
give rise to new stars.
Such structure also mitigates the radiative
transfer in a galaxy, shielding star forming clouds from photodissociation,
or in its absence, leaving clouds mostly dissociated, as dense molecular
cores inside large warm atomic shells. Greenberg was the first to
{\it predict}, almost 30 years ago, the `temperature fluctuations' or
`temperature spiking' of very small dust grains. Again the master's
vision
has been soundly confirmed both by further theoretical and observational
studies.

There is more. Cold (20K and colder) dust grains, acting as a 
cosmic mask or fog, may obscure a
huge $\sim$ 95\% mass fraction of Pop II galactic
backbones. Significant 
dust content can play havoc with attempts to
accurately measure the light and color distributions in a galaxy,
especially if it is embedded  as opposed
to a foreground screen.
Moreover, dust
can play havoc with inferences for the morphological classification
of stellar Population II disks from gaseous Population I speciations,
on which the Hubble tuning fork is based. A entirely new
near-infrared classification scheme emerges when galaxies are {\it
  mask penetrated}; the Hubble tuning fork strikes a new note. 

Greenberg was right on target. Over three decades ago, he
{\it predicted} that 80-90 percent of the dust mass in a galaxy would 
consist of cold and very cold grains (see his pioneering paper `Interstellar
grains and spiral structure' in the 1970 volume edited by H. Habing)
and therefore undetected by IRAS. It is precisely these `large' (tenth
micron) grains which are responsible for the visual extinction in
galaxies; not the smaller, one-hundredth micron grains.

In January 1996, we could all
salute the legendary Greenberg on the fact that optical minus near
infrared imaging
combined with radiative transfer codes, {\it now beautifully
confirmed} his predictions:
dust masses (and therefore dust-gas ratios) can increase by
one order of magnitude. Interarm dust is everywhere. [See the volume
which Mayo and I edited, entitled {\it `New Extragalactic
Perspectives in the New South Africa'}, Kluwer, 1996].

Greenberg's predictions were made some 30 years before
2D direct imaging could routinely become possible longward of I (0.89$\mu m$).
Sub-millimetre/mm observations had brought the issue of cold dust to the
fore, but with controversial and diametrically opposed conclusions.
Quantitative techniques to probe the full spatial extent of dust grains
of all temperatures awaited the commissioning (circa 1990) of large
format HgCdTe NICMOS
arrays
on telescopes at Mauna Kea and elsewhere. We generated
 optical minus near-infrared
colour maps to establish the existence, at arcsecond resolution, of a
cold and very cold population of (interarm) dust,
and Greenberg's predictions
were conclusively verified in 1994 and 1996.

As I reflect over the publication list of Greenberg,
I must, in passing, salute the master in a paper co-authored by
him
in {\it Nature}, concerning the possibility of dust grains in
in galactic
haloes. 4096$\times$4096 pixel CCD imaging of the galaxy
NGC5907 with the Canada France Hawaii Telescope suggests
-- on observational grounds -- that dust
{\it may} be present in its halo.
Colour excesses are at levels of $\sim$ 0.1 magnitude
for the halo NGC 5907, comparable to interarm colour excesses
one finds in dusty galactic disks (eg. NGC4736).
While it remains a great observational challenge to image galactic
haloes, Greenberg blazed the theoretical trail.

Greenberg {\it predicted}

$\bullet$ that the nucleus of Comet Halley
would be black ({\it Nature}, 321, 385, 1986)

$\bullet$ that the cometary dust would be $\sim$ one half
rocky-silicate, the rest being in the form of {\it complex organic
material}.

On both counts Greenberg was right again: predictions confirmed
by the Vega 1, Vega 2 and Giotto missions.

I have only referred to a handful of the implications of the 300+ papers
authored by Greenberg.
In the citation survey conducted by P.C. van der Kruit entitled
`Astronomical
Community in the Netherlands' [published in the {\it Quarterly Journal
of
the Royal Astronomical Society} (vol. 35. no. 4, pg 421, 1994)],
Professor Greenberg
held the highest number of cited publications of any
astronomer in the Netherlands for the year 1991. No mean achievement
for a researcher then at age 70!
We indeed stand on the shoulders on giants.

Thus far, I have only focussed on
some of the {\it publications} of Greenberg.
I have not alluded to the {\it doctoral students} Professor
Greenberg
has mentored; scientists such as L. d'Hendecourt who are now, in their
master's footsteps, playing
crucial roles in our understandings of, for example, polycyclic aromatic
hydrocarbons.  Neither have I alluded to students who have
worked with Greenberg in a postdoctoral capacity, and who later went on
to establish research groups of the highest calibre. I think of
Lou Allamandola, for example, who worked as a  postdoc in
Greenberg's Laboratory at Leiden.

What a fitting way it was to salute the man in 1997, when I invited
astronomers from around the globe to send in their birthday wishes
to Mayo on his 75th
birthday. Scores, dozens, of emails, poured in. George Miley wrote the
first congratulatory email:

{\it ``Your presence during the last two decades has been of immense
benefit to the prestige of Leiden. Your immense enthusiasm and
stimulating attitude to research is a joy to see. I cannot believe
that you are three quarters of a century young. I regard it as a great
privilege to have you as a colleague and friend. On behalf of Hanneke
and myself, Happy Birthday!''}

Next followed an email to my office from Sir Martin Rees,
Astronomer Royal. This message was followed 
by emails from Ron Allen, the Elmegreens,
Y. Terzian, G. Herbig, B. Draine,  A. Li, F. Israel, J. Mather, and about 100 
more researchers.

Walt Duley (University of Waterloo) pleaded:

{\it ``Happy 75th, Mayo! Please don't answer ALL the outstanding questions
on dust before you retire. Leave a few for the rest of us to deal
with! Your friend, Walt Duley.''}

Mayo always remembered his viewing of Comet Halley at the home of 
Bruce and Debbie Elmegreen. Bruce recalled:

{\it ``...you never failed to impress me with the breadth and originality
of your work. I also have the fondest memories of viewing Halley's
comet with you in our back yard. I wish you and your family many many
more happy years. Bruce Elmegreen.''}

George Herbig, in his congratulatory email, included a small, but
 precise,
 calculation:

{\it ``Dear Mayo: May your life continue to be filled with dust,
  molecules, and other assorted interstellar debris! Best regards and
  congratulations on passing the 2.366 $\times$ $10^{9}$ $s$ milestone,
  George Herbig.''}

Kalevi Mattila recalls:

{\it ``Your visits to Finland and your presentations are well remembered
  \\ still today -- not least the magic tricks you presented to my (at
  that time) small children. The first international meeting I
  attended -- the interstellar dust conference in Jena in 1969, was
  organized under your leadership. Ever since then it has always been
  a great pleasure and inspiration for me meeting you at
 different
  places and conferences...''}

Martin Cohen (Berkeley) summed it up well:

{\it ``That yellow soup obviously agrees with you!''}

From founding the microwave scattering laboratory
at the Rensselaer Polytechnic Institute in New York, to the Laboratory of
Astrophysics in Leiden, we continued to see
a man not only characterized by a lifetime of trailblazing research,
but there was more. Like his beloved hero Einstein,
Mayo was, in his chosen field, a 
giant of a man  whose profound {\it predictive} theoretical insights
were truly astounding. As Professor John Kerridge (UCSD) emphasized:

{\it ``You have left an indelible and constructive mark in a host of
  fields, but of course most notably in the relationships between
  interstellar grains, comets, and primitive solar systems. Like so
  many others, I have learnt a great deal from your numerous studies
  in that area...''} 

Mayo was a family man, and Naomi was
his closest earthly companion. The love they shared --  how they still
held hands whilst briskly walking the streets of Leiden --  are
especially etched in my
mind. For me to be included in his innermost
circle of galactic research companions was an immense privilege.
When I first met Mayo, it seemed as if we had known one another for
years. There was a mutual bond -- a tie of {\it extraordinary friendship}
(Figure 5).  

The Greenberg home on the Rhine... 
the hospitality which
Naomi and Mayo extended to my wife Liz and me  at {\it Morsweg} 
over the years can, and
will, never be forgotten. Taxis were `forbidden'. On my last arrival 
at Schipol Airport,
to speak at the Oort Centenary, Mayo and Naomi were both there to meet
me. To quote George Miley: their hospitality was {\it legendary}.

Mayo had an exuberance for Life -- with a captal L. He loved to LIVE. He
loved to {\it encourage}. He visited me in South Africa on two
occassions (figures 6-11), and we also walked the streets of Paris.  
We visited the  Mus{\'e}e Rodin together. He marvelled at Camille
Claudel's onyx and bronze {\it La Vague} and at Rodin's 
{\it The Walking Man, The
  Eternal Idol, La Tour du Travail} and many others. 
Each visit by him was filled with enough dreams to last for a
lifetime.

 When we went game viewing in our South African reserves, Mayo was always there
-- right in the front seat of the 4$\times$4. His eye, always eager to
spot a lion kill; a herd of elephant; a stalking cheetah.... At meal
time, Mayo kept all enthralled. I recall staying at his home about a
year or two before he passed away ... and to see a man, close to 80,
carrying his back-pack, rushing off to catch the train (en route to the
airport) to yet {\it another} Conference abroad, left an indelible
impression on me.

There is a lot I will not say, because they are personal memories. But
let me say this: Mayo enriched my life. He encouraged me, never to give
up. He supported every facet of my research. Figure 12 shows but one
example. We probably spoke on the
telephone at least twice a month -- often much more. If not by phone, we
constantly emailed one another. We thoroughly enjoyed chatting about a
huge range of topics, including the magnificent near-infrared 
spectra of the brown
dwarf Gliese229B by Tom Geballe, B.R. Oppenheimer, S.R. Kulkarni and
their collaborators (Figure 13). 

I had already partly constituted a Scientific Organising Committee to organise
a Conference here in South Africa for Mayo's 80th; much support was
received from George
Miley, Francoise
Combes, Johan Knapen, Duccio Macchetto, Bruce Elmegreen and  Ken
Freeman,
among others. But it was not to be. Fortunately George
Miley, Ewine van
Dishoeck and Willem Schutte were able to quickly organise a 
meeting   
in Mayo's honour, in Leiden; Mayo was physically still strong enough to
attend.  

I  
 spoke to Mayo in his hospital bed in Belgium. We spoke 
just before he passed on, at his home in {\it Morsweg}. 
He was almost too weak to
speak. But there came the voice, weak but {\it still speaking of
future research plans and about dust in the high redshift universe}.

That is the impetus of the work presented here: to explore the
efficiency and possibility of penetrating masks of
cosmic dust at high-z, with NGST.

Mayo was one of my very closest
research companions, and we miss him sorely.

\vskip20pt

{\bf 2004 Greenberg Memorial Public Lecture}

In 2004, we plan to fly out a researcher from abroad 
to South Africa, to
deliver a public lecture at our University Great
Hall (seating capacity, 1000 persons) in honour of J. Mayo Greenberg.
The lecture will probably coincide with our International Conference
{\it Penetrating masks of Cosmic Dust: The Life Duty Cycle of Bars}
which takes place in June 2004. Mayo would indeed have returned
to South Africa, but his health declined so fast. In was in South
Africa that Mayo met the Cabinet Minister for  Arts, Culture, 
Science and Technology, Dr Ben Ngubane (see figure 9). Dr Ngubane is
intimately involved with the 10-m Southern African Large
Telescope SALT  being constructed in South Africa.  

The lecturer need not have personally have worked with Professor
Greenberg, but the theme of the lecture must obviously be one close to
one of Greenberg's passions -- whether it be comets, extrasolar systems, 
the dusty ISM, or the dusty high redshift universe. 
Interested persons are invited to email one of the authors
(at block@cam.wits.ac.za) for further information: it is anticipated
that one international airfare and local accomodation will be fully paid for.

\vskip20pt

  {\bf Acknowledgments}

  DLB expresses his deepest thanks to Iv\^anio Puerari, Marianne
  Takamiya and Bob Abraham, 
  his collaborators on this NGST 922 restframe $K$-band simulation 
  project.

  DLB is indebted to the Anglo American Chairman's
  Fund
  and to Clem Sumter, M. Keeton, Hugh Rix and the Board of
  Trustees. This research was also supported by SASOL; a warm note of
  gratitude
  to Chairman Mr P. Kruger and to the SASOL Board.

  DLB thanks E. van Dishoeck and W. Schutte for the opportunity to write
  his reflections on JMG in their forthcoming publication, too. 
  He is most indebted to George Miley for 
  keeping him fully abreast (both by telephone and email)
  of all developments, as Mayo's days neared their end. 

  DLB recalls Mayo's great excitement at attending Yvonne Pendleton's
  stardust and planetesimals conference in California (subsequently 
  published in the ASP Conference Series), and
  for the purposes of releasing this paper on the WWW, we have compiled 
  this paper using the Latex
  stylefile of the Astronomical Society of the Pacific. 
  The actual stylefile to be used 
  will 
  be
  decided upon
  by the 2004 Cosmic Dust/Bar Conference editors 
  Ken Freeman, David Block, Iv\^anio Puerari and Robert Groess.


\begin{references}
\reference Abraham R.G., van den Bergh, S., Glazebrook, K., Ellis,
R.S., Santiago, B.S., Surma, P., Griifiths, R.E. 1996, \apjs, 107, 1
\reference Bally, J., Morse, J.A. 1999,
{\it American Astronomical Society Meeting}, 195, 86
\reference Baum, W.A. 1972, in External Galaxies and Quasi-Stellar Objects,
Eds. D.S. Evans, D. Wills \& B.J. Wills
(Dordrecht Reidel), 393
\reference Bertin, G., Lin, C.C., Lowe,
S.A., Thurstans, R.P. 1989, \apj, 338, 78
\reference Bertin, G. \& Lin, C.C. 1996, A Density Wave Theory (MIT Press)
\reference Block, D.L. 1996, in New Extragalactic Perspectives in the New South Africa,
Eds. D.L. Block \&  J.M. Greenberg
(Kluwer Dordrecht), 1
\reference Block, D.L., Wainscoat, R.J. 1991, {\it Nature}, 353, 48
\reference Block, D.L., Bertin, G., Stockton, A., Grosb\o l, P., Moorwood, A.F.M.,
Peletier, R.F. 1994, \aap, 288, 365
\reference Block, D.L., Puerari,
I. 1999, \aap, 342, 627
\reference Block, D.L., Puerari, I., Frogel, J.A.,
Eskridge, P.B., Stockton, A., Fuchs, B. 2000, in Toward a New Millennium in
Galaxy Morphology,
Eds. D.L. Block, I. Puerari, A. Stockton \& dW. Ferreira (Kluwer Dordrecht),
5
\reference Block, D.L., Puerari, I., Takamiya, M., Abraham, R., 
Stockton, A., Robson, I. \& Holland, W. 2001, \aap, 371, 393
\reference Block, D.L., Bournaud, F., Combes, F., Puerari, I. \& Buta,
R.J. 2002, \aap, 394, L35
\reference Bournaud, F. \& Combes, F. 2002, \aap, 392, 83
\reference Charlot, S., Worthey, G., Bressan, A. 1996, \apj, 457, 625
\reference Ellis, R.S. 1997, \araa,  35, 389
\reference Elmegreen, D.M., Elmegreen, B.G., 1987, \apj, 314, 3
\reference  Ferguson, H.C.,
Babul, A. 1998, \mnras, 296, 585
\reference Giavalisco, M., Livio, M., Bohlin, R.C., Machetto, F.D., Stecher, T.P. 1996,
\apj, 112, 369
\reference Gillett, F., Mountain, M. 1998, {\it ASP Conference Series}, 133, 42
\reference 
Hauser, M.G. 1994, {\it American Astronomical Society Meeting}, 185,
32
\reference Jarrett, T.H., et al. 2003, AJ, 125, 525 
\reference Kalnajs, A.J. 1975, in
La Dynamique des Galaxies Spirales,
Ed. L. Weliachew (Paris Editions du CNRS),
103
\reference Morgan, W.W. 1958, \pasp, 70, 364
\reference Norman, C.A., Braun, R. 1996, in Cold Gas at High Redshift, Eds.
M.N. Bremer, P.P. van der Werf, H.J.A. R\"ottgering \& C.L. Carilli
(Kluwer Dordrecht), 3
\reference Puerari, I., Block, D.L., Elmegreen, B.G., Frogel, J.A., Eskridge, P.B.
2000, A\&A, 359, 932
\reference Puerari, I., Dottori H.A. 1992, \aaps, 93, 469
\reference Sanders, D.B. 2000, in
Toward a New Millennium in Galaxy Morphology, Eds. D.L. Block, I. Puerari,
A. Stockton \& dW. Ferreira (Kluwer Dordrecht), 381
\reference Takamiya, M. 1999, \apjs, 122, 109
\reference de Vaucouleurs, G., de Vaucouleurs, A., Corwin, H.G., Buta, R.J., Paturel, G.,
Fouqu\'e, P. 1991,
Third Reference Catalogue of Bright Galaxies (Springer Verlag New York)
\reference Williams, R.E. {\it et al.}
1996, \aj, 112, 1335
\reference Windhorst, R., Odewahn, S., Burg, C., Cohen S., Waddington I. 2000, in
Toward a New Millennium in Galaxy Morphology, Eds. D.L. Block,
I. Puerari,
A. Stockton \& dW. Ferreira (Kluwer Dordrecht), 243

\end{references}
  \end{document}